\documentclass{PoS}

\title{Round Table on Axions and Axion-like Particles}

\ShortTitle{Axion}

\author{
{Paolo Di Vecchia}
\\
       The Niels Bohr Institute, University of Copenhagen,\\
       Blegdamsvej 17, DK-2100 Copenhagen, Denmark \\
       and \\
       Nordita, KTH Royal Institute of Technology and Stockholm University\\
       Roslagstullsbacken 23, SE-10691 Stockholm, Sweden \\
        E-mail: \email{divecchi@nbi.dk}}

\author{Maurizio Giannotti\\
Physical Sciences, Barry University,
11300 NE 2nd Ave., Miami Shores, FL 33161, USA\\
E-mail: \email{mgiannotti@barry.edu}}

\author{Massimiliano Lattanzi\\
        Istituto Nazionale di Fisica Nucleare,
        Sezione di Ferrara, Polo Scientifico e Tecnologico, Edificio C,
        Via Giuseppe Saragat, 1, IT-44122 Ferrara, Italy\\
        E-mail: \email{lattanzi@fe.infn.it}}

\author{Axel Lindner\\
Deutsches Elektronen-Synchrotron DESY, 
				Notkestrasse 85, D-22607 Hamburg, Germany\\
        E-mail: \email{axel.lindner@desy.de}}

\abstract{
	In this contribution, based on the discussion at a round table of the XIII Quark Confinement and the Hadron Spectrum - Confinement conference, 
we review the main properties of the QCD axion and, more generally, of axion-like particles and their relevance in astrophysics and cosmology. 
In the last section we describe the experimental concepts to search for the QCD axion and axion-like particles (ALPs).}

\FullConference{XIII Quark Confinement and the Hadron Spectrum - Confinement2018\\
		31 July - 6 August 2018\\
		Maynooth University, Ireland}

\begin{document}

\section{Introduction}

The research on axions is developing very quickly, both at the theoretical and experimental level, and involves expertise  from astrophysics, cosmology and various aspects of particle physics. The aim of the round table was to discuss  the various  aspects of the research on axions and conclude with an open discussion between us and the audience. We start here with an introduction to the problem.

The original motivation for axions comes from QCD.  In fact,  to the usual QCD action, one can add a $\theta$ term 
\begin{eqnarray}
L =  -\frac{1}{4} F_{\mu \nu}^a F^{a \mu \nu}  +i {\bar{\Psi}} \gamma^\mu D_\mu \Psi - {\bar{\Psi}} M \Psi  -  \theta   \,\,Q (x) \,\,\, ,
\label{A1}
\end{eqnarray}
where $Q$ is the topological charge density:
\begin{eqnarray}
Q (x) = \frac{1}{32 \pi^2} F_{\mu \nu}^a {\tilde{F}}^{a \mu \nu}~~;~~ {\tilde{F}}_{\mu \nu}^a = \frac{1}{2} \epsilon_{\mu \nu \rho \sigma} F^{a \rho \sigma}  \,\, .
\label{A2}
\end{eqnarray}
The $\theta$ term, together with a phase in the quark mass matrix,  breaks $CP$ and produces a non-zero electric dipole moment for the neutron~\cite{Crewther:1979pi} (in units of $e=1$):
\begin{eqnarray}
D_n  \sim   {\bar{\theta}} \cdot 3.6 \cdot 10^{-16} \,\mathrm{cm} ~~;~~
{\bar{\theta}} = \theta + Arg \det M \,\, .
\label{A3}
\end{eqnarray}
The experimental limit~\cite{Afach:2015sja} 
\begin{eqnarray}
|D_n| < 3.0 \cdot 10^{-26} \,\mathrm{cm}  \Longrightarrow {\bar{\theta}} \lesssim 10^{-10}
\label{A4}
\end{eqnarray}
gives a value for ${\bar{\theta}}$ that is very small and actually consistent with  zero.
Can we make it to be zero in a natural way?  

The proposal of Peccei and Quinn~\cite{PQ}, for making it to be zero, has been to  introduce, in the matter sector of QCD, some new degree of freedom with an extra $U(1)_{PQ}$ symmetry that is broken by an anomaly exactly as the $U(1)_A$  of QCD. It has been then realized that the Peccei-Quinn mechanism implies the existence~\cite{SWax,FW,BT} of a new particle that was called axion.
For reviews on axions see Refs.~\cite{KIM,Peccei,RAF,PS}.

A convenient way to describe its connection with the low energy degrees of freedom of QCD is to include~\cite{DVS,DiVecchia:2017xpu} the axion in the effective Lagrangian~\cite{RST,DVV,NA,EW,KO,NO,PDV} for the  pseudo-scalar mesons:
\[
L = \frac{1}{2} Tr ( \partial_{\mu} U \partial_{\mu} U^{\dagger} ) + 
\frac{1}{2}  \partial_{\mu} N \partial_{\mu} N^{\dagger}  +
\frac{F_{\pi} }{2 \sqrt{2}} Tr \left( \mu^2 ( U + U^{\dagger} ) \right) 
 +
\]
\begin{eqnarray}
- \theta Q + \frac{Q^2}{a F_{\pi}^{2}} +
\frac{i}{2} Q(x) \left(Tr (\log U - \log U^{\dagger})  + \alpha_{PQ}(\log N - \log
  N^{\dagger} ) \right) 
\label{A5}
\end{eqnarray} 
where $U$ is the field describing the pseudo-scalar mesons, while $N$ is the one related to the axion $a$
\begin{eqnarray}
U (x) = \frac{F_{\pi}}{\sqrt{2}} e^{ i \sqrt{2} \Phi
  (x)/F_{\pi}}~;~
N (x) = \frac{F_{a}}{\sqrt{2}} e^{ i \sqrt{2} a
  (x)/F_{a}}~;~ \mu^2_{ij} = \mu^2_i \delta_{ij} \,\, .
\label{A6}
\end{eqnarray} 
$\mu^2_{ij}$ is related to the quark mass matrix that gives a non-zero mass to both the axion and the pseudo-scalar mesons  and that mixes  them. In general, however,  the axion is also coupled to all the fields of the Standard Model   through the following  effective Lagrangian
 (see for instance Ref.~\cite{GV} for a study of the properties of the  QCD axion):
\begin{eqnarray}
L_{axion} =  \frac{1}{2} \partial_\mu a \partial^\mu a - \frac{\alpha_{PQ} \sqrt{2}a}{F_a}  Q(x) + \frac{g_{a \gamma \gamma}}{4} F_{(em)} \cdot  {\tilde{F}}_{(em)} a +  L_{int} \left( \frac{\partial_\mu a}{F_a} , \Psi \right)
\,\,\, .
\label{A7}
\end{eqnarray}
The first two terms are universal and are the same as the second and the last terms  appearing in Eq. (\ref{A5}). The third term describes the coupling of the axion to two photons where 
\begin{eqnarray}
 g_{a\gamma \gamma} = \frac{\alpha\sqrt{2}}{2\pi F_a} \left( \frac{E}{N} - \frac{2}{3} 
\frac{m_u +4 m_d}{m_u +m_d}  \right)~~;~~ \frac{m_u}{m_d} \sim 0.5 \,\, .
\label{A8}
\end{eqnarray}
$E$ and $N$ are respectively the electromagnetic and the colour anomaly of the axial current associated to the axion field. In particular,  $\frac{E}{N} = \frac{8}{3}$ in grand-unified models as in
the DFSZ model~\cite{ARZ,DFS} and $\frac{E}{N} =0$ in the KSVZ model~\cite{JKIM,SVZ}. 
Finally, the last term in Eq. (\ref{A7}) is also not universal.

From the Lagrangian in Eq. (\ref{A5}) it can be seen that the axion gets a non-zero vacuum-expectation-value  that cancels the dependence of the physical quantities on $\theta$. It can also be seen that the axion mixes with the neutral pseudo-scalar mesons and gets a non-zero mass given by
\begin{eqnarray}
m_a^2 =   
\frac{2\alpha_{PQ}^2}{F_a^2} \chi_{QCD}
~~~~;~~~ \chi_{QCD} = \langle Q(0) \int d^4 x \,\, Q (x) \rangle \,\, ,
\label{A9}
\end{eqnarray}
where $\chi_{QCD}$ is the topological susceptibility in QCD.

The mechanism of Peccei-Quinn with the introduction of a $U(1)_{PQ}$ symmetry that is then broken by the $U(1)_A$ anomaly may look very artificial. On the other hand, many of this kind of $U(1)$'s appear in string theory~\cite{SW}. 

The $10$-dimensional  consistent string theories contain higher  anti-symmetric gauge potentials  ($B_{\mu \nu} , C_{\mu \nu \rho} \dots$) that generalize the electromagnetic  potential $A_\mu$.
Phenomenologically viable models require to compactify the 6 extra dimensions: $M_{10} = M_4 \times V_6$. Therefore, for each choice of compactification, we get a  four-dimensional  Lagrangian  containing, in general,  several four-dimensional  $B_{\mu \nu}$ potentials. In $D=4$ they correspond to pseudo-scalars fields $A$ through  the duality relation:
\begin{eqnarray}
\epsilon_{\mu \nu \rho \sigma} \partial^\nu B^{\rho \sigma} \sim  \partial_\mu A
\label{A10}
\end{eqnarray}
with Lagrangian (as the QCD axion)
\begin{eqnarray}
L = \frac{1}{2} \partial_\mu A \partial^\mu A + \frac{A}{F_A} Q(x) + \dots ~;~F_A \sim \frac{\alpha_{G} M_P}{2\pi \sqrt{2}} \sim 10^{16} \,\mathrm{GeV}~;~ \alpha_G = \frac{1}{25}
\label{A11}
\end{eqnarray}
where $M_P$ is the Planck mass and $\alpha_G$ the fine structure constant in grand-unified theories.
The study of Ref.~\cite{SW} indicates that, in general, there is the tendency to get too large values for $F_A$ close to $M_P$. The important thing is, however, that those pseudo-scalars, unlike their scalar partners that are required to get masses of  the order of $10$ TeV to avoid problems with fifth forces, can have a very small mass at the sub eV level. See  for instance Ref.~\cite{CGR} for a detailed discussion of these questions in the framework of compactifications on Calabi-Yau spaces and Ref.~\cite{JR} for a review.  

In this introduction we have briefly summarized some property of the QCD axion and of axion-like particles (ALPs). In the following we will be describing various aspects of the physics of the QCD axion and of  ALPs.

\begin{figure}[h]
	\begin{center}
		\includegraphics[width=0.65\linewidth]{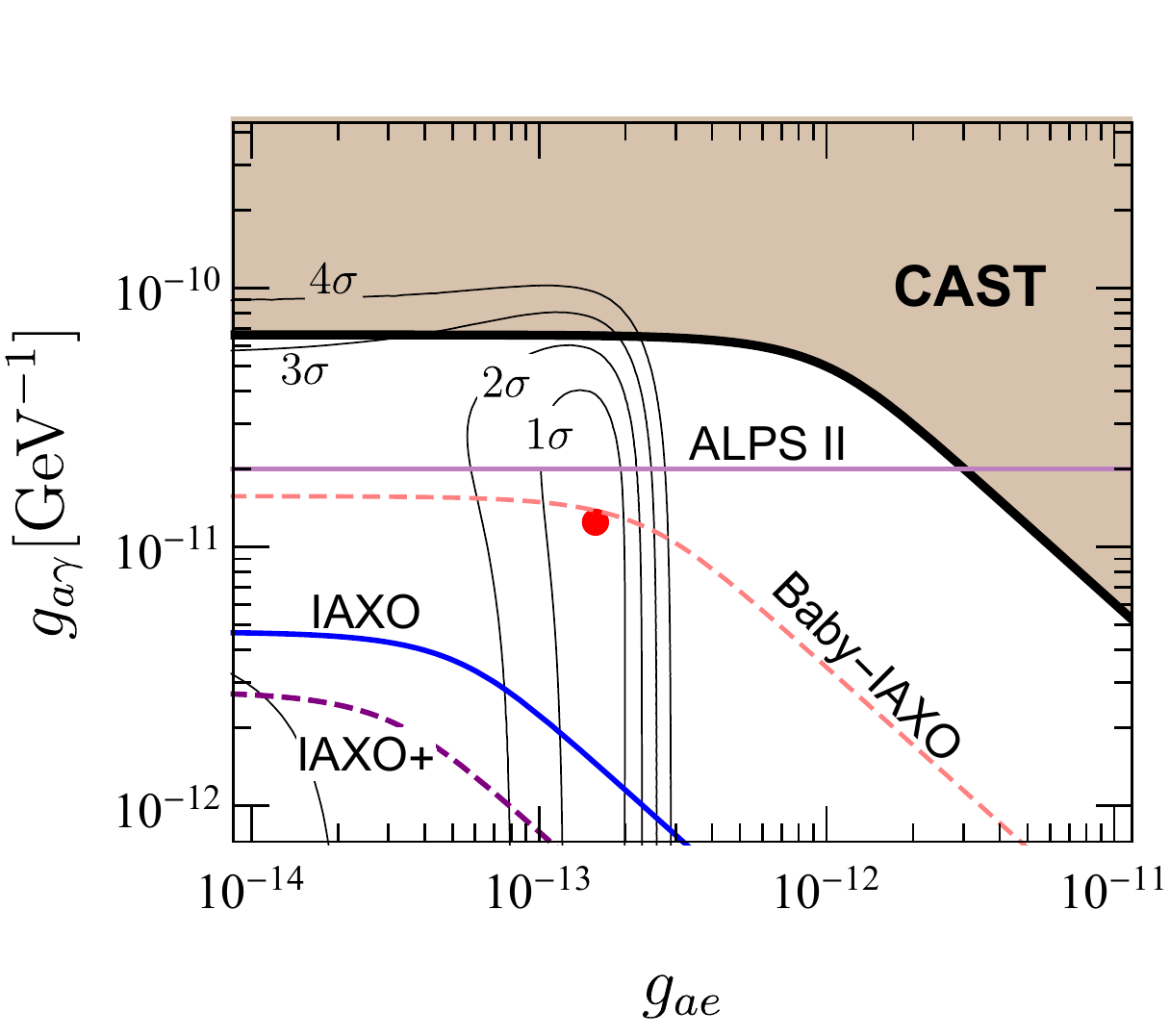}
		\caption{Hinted regions in the ALP parameter space from stellar observations.
			The experimental potential is also shown. 	}
		\label{fig:sensitivity}
	\end{center}
\end{figure}

\section{Axions in astrophysics}

\begin{figure}[t]
	\begin{center}
		\includegraphics[width=0.65\linewidth]{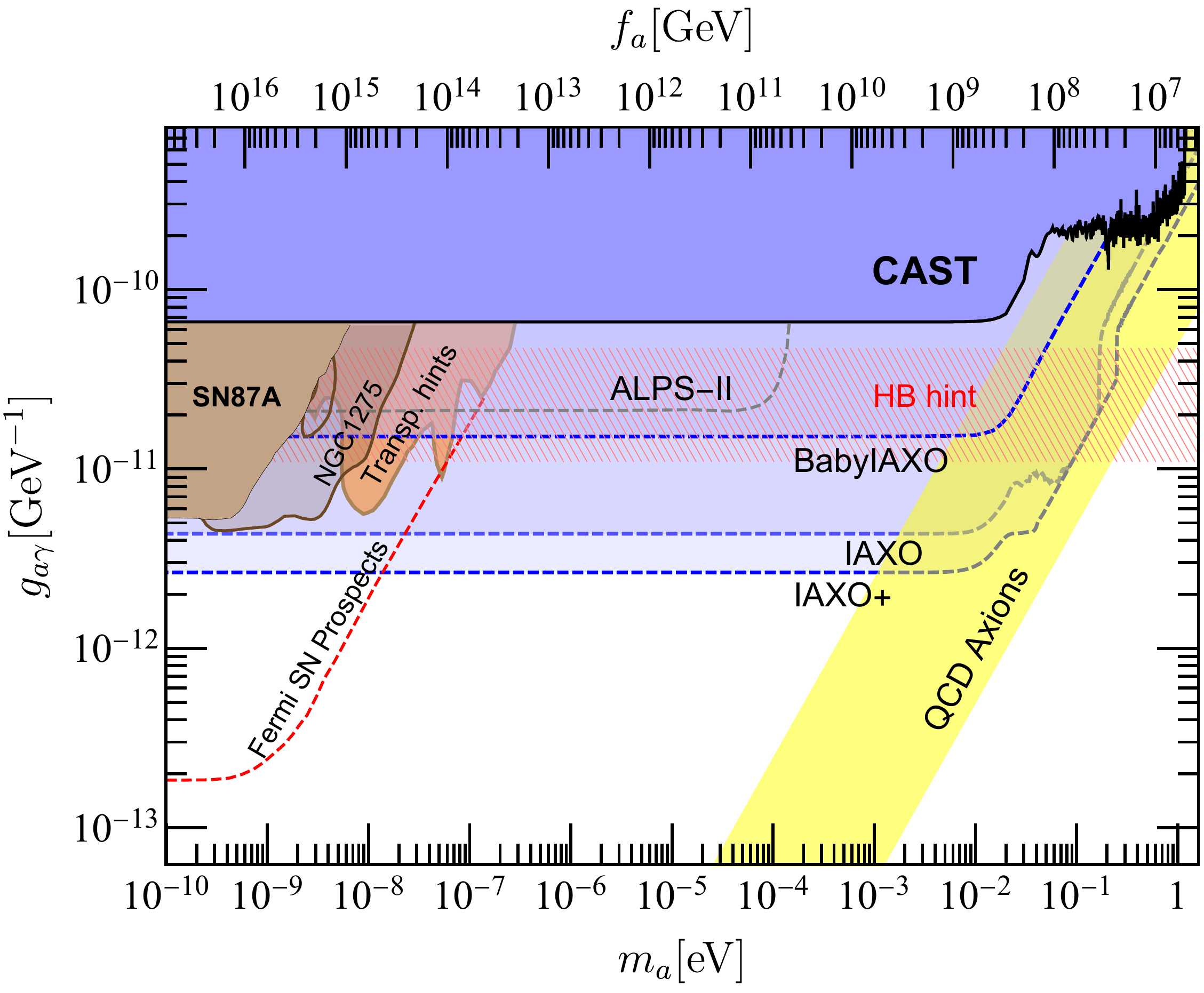}
		\caption{ALP parameter space $ g_{a\gamma} $ vs. $ m_a $ with axion hints and experiments. 
			The region hinted by HB stars is calculated for ALPs interacting only with photons. 
			}
		\label{fig:gag_ma}
	\end{center}
\end{figure}

Besides dedicated terrestrial experiments, stellar observations offer a unique - and often very powerful - way to look at axions and other weakly interacting particles~\cite{RAF,Raffelt:1996wa}. 
Considerations about stellar evolution have provided very strong bounds on the axion couplings to photons, electrons, and nucleons, often exceeding the results achieved in laboratory experiments. 

Quite intriguingly, a series of astrophysical observations have shown an excessive energy loss in many stellar systems, which could be accounted for by additional light, weakly interactive particles. 
These \emph{anomalous} observations include \textit{i)} several pulsating white dwarfs (WDs), in which the cooling efficiency was extracted from the rate of the period change~\cite{Isern:1992gia,Corsico:2012ki,Corsico:2012sh,Corsico:2016okh,Battich:2016htm}; 
\textit{ii)} the WD luminosity function (WDLF), which describes the distribution of WDs as a function of their brightness~\cite{Isern:2008nt,Bertolami:2014wua,Isern:2018uce};
\textit{iii)} red giants branch (RGB) stars, in particular the luminosity of the tip of the branch~\cite{Viaux:2013lha,Straniero:2018fbv}; 
\textit{iv)} horizontal branch stars (HB) or, more precisely, the R-parameter, that is the ratio of the number of HB over RGB stars~\cite{Ayala:2014pea, Straniero:2015nvc};
\textit{v)} the ratio of blue and red supergiants~\cite{McQuinn:2011bb,Skillman:2002aa,Friedland:2012hj,Carosi:2013rla}.

The \textit{new-physics} interpretation of these anomalies has resulted in the selection of axions and ALPs, among the various light, weakly interacting particles, as the only candidates that can explain all the excesses~\cite{Giannotti:2015kwo}.
Given the very different stellar systems in which excessive energy losses have been observed, it is quite remarkable that one single candidate can explain all the observations.
A global analysis of the hints from WDs, HB, and RGB stars, shown in Fig.~\ref{fig:sensitivity}, indicates a preference for ALPs coupled to both electrons and photons, though a vanishing photon coupling would still be compatible with the observations within 1$ \sigma $.
The electron coupling, on the other hand, is predicted to be finite with a $ \sim 3\sigma $ statistical significance.
In any case, the hints point to a well defined area in the axion parameter space which is in part accessible to the next generation of axion probes (see sec.~\ref{sec:experiments} for more details).

Specific QCD axion models could also be responsible for the excessive cooling~\cite{Giannotti:2017hny}.
In this case, there are well defined relations between the different axion couplings with the standard model fields and, in particular, the axion couplings to nucleons cannot be neglected. 
Strong bounds on the axion-nucleon couplings were inferred from the observed neutrino signal of SN 1987A and, more recently, from the cooling of the neutron stars in the supernova remnant Cassiopeia A~\cite{Hamaguchi:2018oqw} and the neutron star in HESS J1731-347~\cite{Beznogov:2018fda}.
If we take the DFSZ axion model as an example, we find that the combined observations point to the mass range $ \sim (2 - 20) $ meV~\cite{Giannotti:2017hny}. 
This mass range is difficult to probe experimentally (see Fig.~\ref{fig:gag_ma}), and partially accessible only to IAXO and, possibly, ARIADNE~\cite{Arvanitaki:2014dfa}.

These bounds on the nuclear couplings are, however, quite less robust than the bounds from other stars, in part because of the difficulty in describing the nuclei interactions in the axion bremsstrahlung process (see, e.g.,~\cite{Chang:2018rso}).
Moreover, it is possible to consider specific nucleophobic axion models~\cite{DiLuzio:2017ogq}, in which the tension with these bounds could be strongly relaxed.
In this case, the hinted mass region would move to about 0.1 eV~\cite{Giannotti:2017hny}, a fairly large value, possibly accessible only to IAXO among the proposed next generation of axion-scopes.
Interestingly, this region has also been proposed recently to explain the EDGES observation of the anomalously strong 21cm absorption feature~\cite{Houston:2018vrf,Houston:2018vbk}.   

Another longstanding astrophysical puzzle, the excessive transparency of the universe to high energy (E $ \gtrsim $ 100 GeV) photons in the galactic and extragalactic medium~\cite{Protheroe:2000hp}, has also invoked an ALP solution~\cite{Horns:2012fx,Meyer:2013pny,Rubtsov:2014uga,Korochkin:2018tll}.
High energy photons produce electron-positron pairs when scatter on the extragalactic background light.
Therefore, the Universe should be opaque to high energy photons, with the threshold energy depending on the source distance,
and the number of very high energy gamma rays from extragalactic sources should be strongly attenuated. 
However, a number of sources have been observed at large optical depths and have shown unexpected hardening at high energies.
Several studies, culminated in the updated analysis in~\cite{Korochkin:2018tll}, which was
performed on an enlarged and high-quality sample of sources, confirmed the anomalous hardening and identified a light ALP coupled to photons as a very elegant solution.

In the ALP solution to the transparency problem, photons are converted into ALPs in the extragalactic magnetic field~\cite{Raffelt:1987im} and, after propagating unimpeded for some distance, are reconverted into photons in a mechanism largely resembling the light shining through a wall experimental setup.
In this case, the observations hint to a much lower mass region, excluding the possibility of a QCD axion solution.
Nevertheless, the hinted range of the axion-photon coupling largely superimposes the range hinted by the stellar cooling excesses (see Fig.~\ref{fig:gag_ma}).

Part of the hinted region has been excluded by the non-observation of gamma rays from SN~1987A~\cite{Payez:2014xsa} and, more recently, by the search for spectral irregularities in the gamma ray spectrum of NGC~1275~\cite{TheFermi-LAT:2016zue}.
However, a large section remains available.
As evident from Fig.~\ref{fig:gag_ma}, this region is accessible to next generation of axion experiments such as ALPS II and IAXO.
Also, remarkably, the entire region hinted by both cooling anomalies and transparency would be easily accessible to the Fermi Large Area Telescope in the event of a future galactic SN~\cite{Meyer:2016wrm}. 

Very recently further evidence for photon-ALP oscillations has been proposed to explain the spectral modulations of galactic pulsars~\cite{Majumdar:2018sbv} and the spectra of supernova remnants~\cite{Xia:2018xbt}. The two independent analyses point to remarkably similar axion parameters, $m_a=$ a few neV and $g_{a\gamma}$= a few $10^{-10}$ GeV$^{-1}$. However the hinted region is in tension with the CAST results from the search of solar axions (cfr. sec.~\ref{sec:experiments} and Fig.~\ref{fig:gag_ma}).

Whether the astrophysical observations are correctly hinting at physics beyond the standard model or not can be verified only with dedicated terrestrial experiments. 
The advances in astrophysical observations expected in the next few years, however, will certainly improve the quality of the data and either reduce or strengthen the significance of these hints.
Whatever the case, it is certainly remarkable that astrophysical analyses and terrestrial experiments of the next generation will be able to explore largely overlapping regions of the ALP parameter space,
a fact never witnessed before. 
This undoubtedly adds to the current appeal of the research on axions. 

\section{Axions in cosmology}

Axions and ALPs might also play a role in the cosmological evolution. Cosmological observations, including measurements of the cosmic microwave background (CMB) anisotropies and of the distribution of large scale structures (LSS), can thus be used to constrain axion properties, providing complementary information to that obtained from laboratory experiments and stellar observations. Interestingly enough, ALPs are possibly related to different aspects of the
cosmological phenomenology, including dark matter, dark energy, dark radiation, and inflation.

A cosmological population of axions and ALPs can be produced through several mechanisms: thermal production, decay of topological defects (e.g. strings), decay of heavy particles, and the misalignment mechanism. Cosmological axions can be either ``cold'' or ``hot''. In the former case they might make up for all, or a large part of, the dark matter content of the Universe. In the latter case, considerations of structure formation require that axions can only be a subdominant component of the dark matter. 
Axions can be produced from the thermal bath, e.g. QCD axions are produced through pion-pion scattering, $\pi + \pi \to \pi + a$ (\cite{KolbTurner,Sikivie:2006ni}; see also \cite{Salvio:2013iaa} for a recent calculation). Being relativistic, these axions are hot and would contribute as dark radiation at early times. The decay of a heavy particle also usually produces relativistic axions, and thus dark radiation. In fact, the presence of an axion population from moduli decay is a generic prediction of string and M-theory (see e.g. \cite{CGR,JR}).
Measurements of the effective number of relativistic species $N_\mathrm{eff}$ can be used to constrain relativistic axions. Present observations are consistent with no radiation components in addition to the three neutrino families of the SM; for example, Planck 2018 data yield $N_\mathrm{eff} = 2.99 \pm  0.34$ (95\% CL)   \cite{Aghanim:2018eyx}. Thermal axions also affect structure formation at late times, suppressing small-scale fluctuations in a similar way to neutrinos. Together, these two effects can be used to constrain the abundance of relativistic axions and thus their mass. A combination of Planck 2015 data (including Sunyaev-Zeldovich cluster counts) with measurements of the Hubble constant and of the baryon acoustic oscillations scale yields $m_a < 0.54 \mathrm{eV}$ at 95\% CL for thermal QCD axions (\cite{DiValentino:2015wba}; see also \cite{Archidiacono:2013cha,Giusarma:2014zza} for older constraints using Planck 2013 data). Bounds on thermal axions can be evaded by relaxing the assumption of a standard thermal history, like in scenarios with a low reheating temperature \cite{Grin:2007yg}, but are quite robust with respect to assumptions about the underlying inflationary model~\cite{DiValentino:2015zta,DiValentino:2016ikp}.
Future observations are expected to improve the sensitivity on $N_\mathrm{eff}$ by one order of magnitude, and might be able to detect $m_a \simeq 0.15 \,\mathrm{eV}$ at high significance, from a combination 
of CMB and LSS observations \cite{Archidiacono:2015mda}. 

Cold axions can instead be produced by the decay of topological defects (strings and domain walls) produced at the time of PQ symmetry breaking \cite{Davis:1985pt}, or by the misalignment mechanism, in which the axion field is initially displaced from its minimum \cite{Preskill:1982cy,Abbott:1982af,Dine:1982ah,Turner:1985si}. Both these mechanisms produce non-relativistic axions, that are thus good candidates for cold dark matter~\cite{KolbTurner,Sikivie:2006ni}. In the misalignment mechanism, the axion field at early times (as long as $H>m_a$) is frozen at its initial value\footnote{Note that here $a$ is the axion field and not the cosmic scale factor.} $a(t_\mathrm{in})$ and behaves as vacuum energy. Later, when $H<m_a$,  coherent oscillations of the axion field around the minimum set up (``vacuum realignment'') and axions behave as non-relativistic matter. The present energy density $\Omega_{\mathrm{mis}}$ of misalignment-produced axions depends on the axion mass, including its evolution with time, and on the initial misalignment angle $\theta_\mathrm{in} = a(t_\mathrm{in})/F_a$. Axion production 
in the decay of cosmic strings is dominated by low-frequency modes, making them also non-relativistic. The relic density of string-produced axions $\Omega_{\mathrm{dec}}$ can be computed through field-theoretic lattice simulations of the string-wall network, and is usually expressed through the ratio $\alpha_\mathrm{dec}\equiv \Omega_\mathrm{dec}/\Omega_\mathrm{mis}$. At the current time, there is no consensus about the value of $\alpha_\mathrm{dec}$, with some studies concluding that the contribution from topological defects (including also domain walls) is dominant over the misalignment mechanisms (i.e., $\alpha_\mathrm{dec} \gg 1$) \cite{Battye:1994au,Wantz:2009it,Hiramatsu:2010yu,Hiramatsu:2012gg,Hiramatsu:2012sc}, and others coming to the opposite conclusion ($\alpha_\mathrm{dec} \lesssim 1$) \cite{Harari:1987ht,Chang:1998tb,Hagmann:2000ja,Kawasaki:2014sqa,Klaer:2017ond}. In any case, individual studies are also affected by significant theoretical uncertainties related to the poor knowledge of some parameters, like e.g. the scale parameter, that characterises the average number of strings per horizon volume. Recently, a logarithmic grow of the scale parameter with asymptotically large times has been reported by different groups \cite{Fleury:2015aca,Gorghetto:2018myk,Kawasaki:2018bzv}. If confirmed, such a growth might enhance the number of axions emitted by strings by a factor of a few \cite{Kawasaki:2018bzv} or even 10 \cite{Gorghetto:2018myk}, meaning that the string contribution to the axion density has been underestimated in previous studies. Further studies are definitely necessary to settle this issue. 
Ref.~\cite{Gorghetto:2018myk} also notes that substantial uncertainties are related to the instantaneous emission spectrum, especially for what concerns the extrapolation of its behaviour to late times.
When discussing vacuum misalignment and string decay as mechanisms for axion production, it is important to distinguish two different cases: the ``pre-inflationary axion`` scenario, in which the PQ symmetry is broken during inflation ($F_a > H_I/2\pi$, with $H_I$ the Hubble parameter during inflation), and the ``post-inflationary axion" scenario ($F_a < H_I/2\pi$), in which it is broken after inflation. Topological defects are created at the time of PQ symmetry breaking, but if this happens during inflation they are inflated away. Thus axion production from topological defects is only relevant in the post-inflationary axion scenario. Moreover, when the PQ symmetry is broken, the axion field in causally disconnected regions acquires a different value for the initial misalignment angle $\theta_\mathrm{in}$. If this happens during inflation, our present observable Universe is contained within a single patch of constant $\theta_\mathrm{in}$, that is then a free parameter of the model. If the PQ symmetry breaking happens after inflation, the observable Universe contains many patches with different values of $\theta_\mathrm{in}$, averaging to a background value $\langle \theta^2_\mathrm{in} \rangle = \pi^2/3$. 

\begin{figure}[h]
	\begin{center}
		\includegraphics[width=0.65\linewidth]{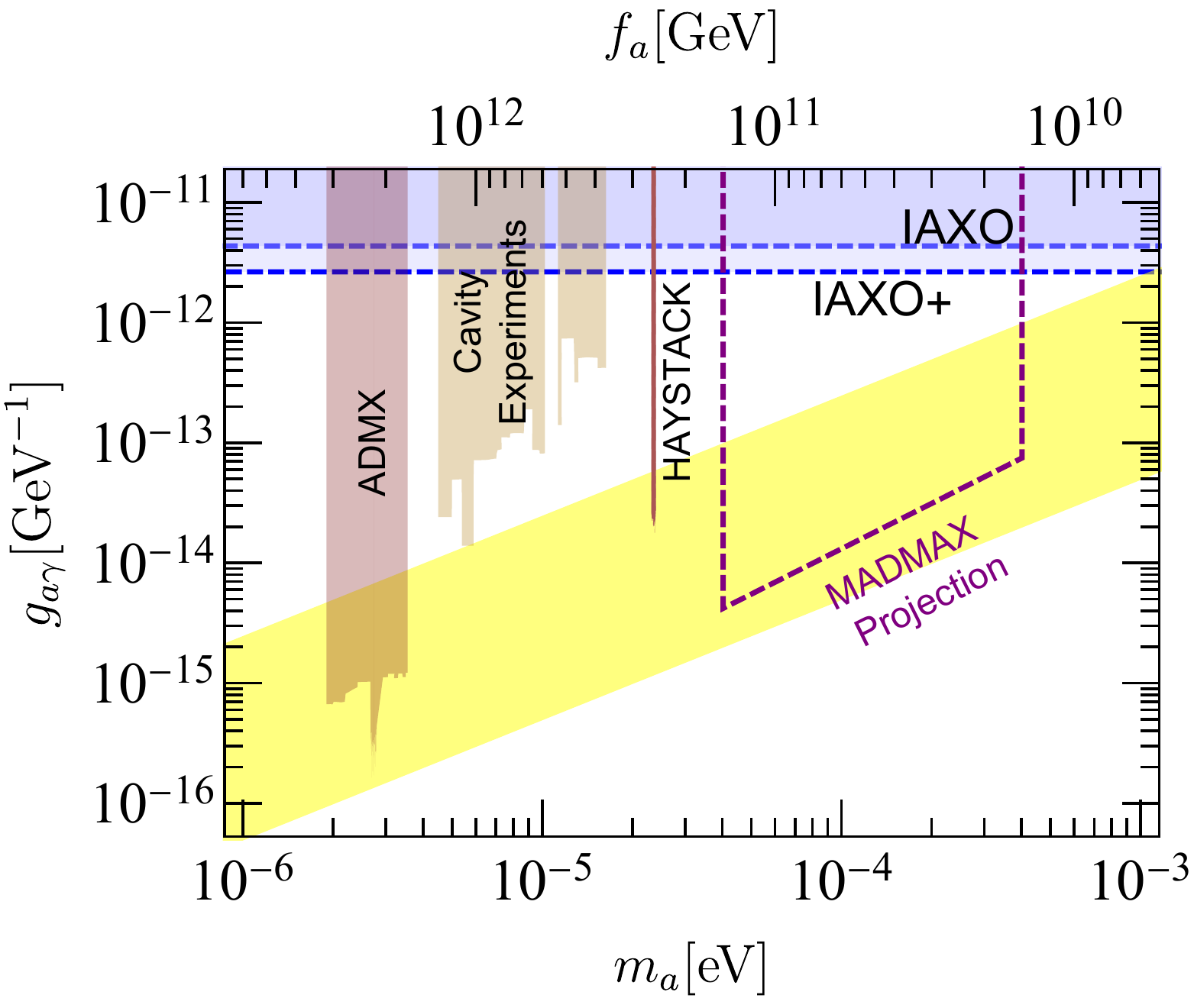}
		\caption{Haloscopes results and projected sensitivities.	}
		\label{fig:haloscopes}
	\end{center}
\end{figure}

In the case of the QCD axion, the axion mass $m_a$  and decay constant $F_a$ are related by the (temperature-dependent) topological susceptibility $\chi_\mathrm{QCD}$ through Eq.~(\ref{A9}). The temperature dependence of $\chi_\mathrm{QCD}$ enters in the determination of the temperature at which the coherent oscillations of the axion field begins, and is needed to evaluate the abundance $\Omega_\mathrm{mis}$ of QCD axions produced through vacuum realignment. The topological susceptibility can be computed using lattice QCD simulations (see \cite{Berkowitz:2015aua,Kitano:2015fla,Borsanyi:2015cka,Bonati:2015vqz,Petreczky:2016vrs,Frison:2016vuc,Borsanyi:2016ksw,Burger:2018fvb,Bonati:2018blm,Giusti:2018cmp} for recent efforts). A useful fit for the misalignment axion density parameter today in the pre-inflationary scenario is \cite{Ballesteros:2016xej} $\Omega_\mathrm{mis} h^2 \simeq 0.12 \left[F_a /(9\times 10^{11}\,\mathrm{GeV})\right]^{1.165} f(\theta_\mathrm{in}) \theta_\mathrm{in}^2$, where $f(\theta_\mathrm{in})$ is a factor that takes into account anharmonicities in the axion potential ($f \to 1$ as $\theta_\mathrm{in}\to 0$), 
and the exponent $1.165$ comes from the temperature dependence of $\chi_\mathrm{QCD}$. Thus, in the pre-inflationary axion scenario, the PQ scale has to be smaller than $\sim 10^{12} \,\mathrm{GeV}$ (or equivalently $m_a \gtrsim 6\,\mu\mathrm{eV}$) if $f(\theta_\mathrm{in})\theta_\mathrm{in}^2\sim 1$, in order not  to  exceed the observed cold dark matter density $\Omega_c h^2= 0.1200 \pm 0.0012$ \cite{Aghanim:2018eyx}.
However, larger values of the PQ scale and smaller axion masses are possible if one allows for a certain degree of fine tuning, in the form of a suitably small value of $\theta_\mathrm{in}$. The case $F_a\simeq 10^{12}\,\mathrm{GeV}$ is referred to as the ``natural'' axion, while the region $F_a \gg 10^{12}\,\mathrm{GeV}$, that requires $\theta_\mathrm{in} \ll 1$, is the so-called ``anthropic'' window. The natural region of axion parameter space can be probed by ADMX (see~Fig.\ref{fig:haloscopes}). In the post-inflationary scenario, instead, the region that evolved to become our observable Universe  contains many patches with different values of $\theta_\mathrm{in}$, and the quantity $f(\theta_\mathrm{in})\theta_\mathrm{in}^2$ should be averaged over the range $\theta\in [-\pi,\,\pi ]$. This yields $\Omega_\mathrm{mis} h^2 \simeq 0.12 \left[F_a /(1.9\times 10^{11}\,\mathrm{GeV})\right]^{1.165}$, so that the right dark matter density is obtained for $F_a\simeq 2 \times 10^{11}\,\mathrm{GeV}$, or $m_a \simeq 30 \,\mu\mathrm{eV}$; smaller values of $F_a$ (larger masses) are thus in principle allowed by cosmology. This region, bounded below by the constraints from stellar physics defines the ``classic'' axion window and will be probed in the next years by MadMAX and IAXO. In this case, however, one should also take into account the production of axions in the decay of topological defects. As noted above, significant uncertainties are associated to this production mechanism. The predictions for the value of $m_a$ that yields the observed DM abundance, once both vacuum realignment and string decay are considered, vary in the range $m_a \approx (25 - 200) \,\mu\mathrm{eV}$ (see e.g. \cite{Wantz:2009it,Klaer:2017ond,Visinelli:2009zm,Visinelli:2014twa,DiValentino:2014zna}). 

If the PQ symmetry is broken during inflation, the presence of another scalar field, in addition to the inflaton, leads to the generation of primordial uncorrelated isocurvature fluctuations, whose power $P_\mathrm{iso}$ is proportional to the combination $H_I^2/\theta_\mathrm{in}^2F_a^2$. Planck observations of the CMB anisotropies constrain uncorrelated isocurvature perturbations to contribute at
most $\simeq 4\%$ of the total primordial power \cite{Akrami:2018odb}, thus providing important information about axions during inflation. 
Assuming that axions make up for all the dark matter, the initial misalignment angle can be expressed as a function of $F_a$,
and the non-observation of isocurvature modes yields a constrain $H_I < 0.86 \times 10^7\,\mathrm{GeV} (F_a/10^{11}\,\mathrm{GeV})^{0.408}$ \cite{Akrami:2018odb}. Then, if we were to know the scale of inflation, this would set a lower limit on the axion decay constant. The energy scale $H_I$ of inflation is currently constrained by the non-observations of tensor modes in the CMB: $H_I<2.7\,M_p \simeq 6.6 \times 10^{13}\,\mathrm{GeV}$ combining Planck 2018 and BICEP2/Keck Array data \cite{Akrami:2018odb}.
An observation of tensor modes in next-generation CMB experiments would imply $H_I \sim 10^{13}\,\mathrm{GeV}$ and require a super-planckian $F_a$ to accomodate the isocurvature bound, thus strongly disfavouring dark matter entirely made by pre-inflationary axions (see also \cite{Visinelli:2009zm,Marsh:2014qoa} for the implications of the, now disproved, BICEP2 claim).

An interesting phenomenological consequence of the post-inflationary scenario is the possibility of forming axion miniclusters \cite{Kolb:1993zz}, that evolve from large spatial variations of the axion field and can host a large fraction of the cosmological axion population.
The density field that will eventually evolve to form axion miniclusters has been recently simulated, also taking into account the contribution from strings and domain walls, in Ref~\cite{Vaquero:2018tib}. Gravitational lensing can constrain axion miniclusters  \cite{Kolb:1995bu,Zurek:2006sy,Fairbairn:2017dmf}.

ALPs can also be produced through the misalignment mechanism. In particular, ALPs with $F_a$ at the GUT scale can provide the observed DM abundance if $m_a \simeq 10^{-19}\,\mathrm{GeV}$ without fine-tuning ($\theta_\mathrm{in} \simeq 1$). Such a value of the mass falls in the region of ultra-light axions (ULAs) (\cite{Hui:2016ltb}; see also \cite{Marsh:2015xka} for a review), defined by the requirement that their Compton wavelength exceeds the Earth radius, or $m_a < 2\times 10^{-13}\,\mathrm{eV}$. Thus for ULAs the particle description breaks down at scales larger than the Earth size (at least). As a consequence, ULAs behave as ``fuzzy'' dark matter, resulting in a different pattern of structure formation at small scales with respect to standard CDM, and might have the potential to solve the ``CDM small-scale crisis'' \cite{Press:1989id,Hu:2000ke}  . ULAs are predicted, for example, in string theory compactifications (see Sec. 1).
In order for the ULAs to contribute to the DM density, coherent oscillations of the ULA field should start before matter-radiation equality; this amounts to the requirement $m_a \gtrsim 10^{-27} \,\mathrm{eV}$. CMB observations constrain $m_a > 10^{-24}\,\mathrm{eV}$ for ULAs making up all the DM \cite{Hlozek:2014lca}. Stronger constraints can be obtained from non-linear probes like the Ly-$\alpha$ forest flux power spectrum, see e.g. \cite{Irsic:2017yje}. ULAs lighter than the present-day Hubble constant, $m_a \lesssim H_0 \sim 10^{-33}\,\mathrm{eV}$ are instead still ``frozen'' at the initial misalignment angle and thus behave as a cosmological constant. In the region of masses between $10^{-32}\,\mathrm{eV}$ and $10^{-26}\,\mathrm{eV}$, ULAs can contribute at most $5\%$ of the total DM density \cite{Hlozek:2014lca}.

Axions, or better ALPs, can also drive inflation\footnote{This is different from the pre-inflationary scenario discussed previously, in which the axion is not the inflaton but only a spectator field during inflation.}.
The simplest inflationary scenario involving the axion is the so-called ``natural inflation''  \cite{Freese:1990rb}, in which the inflationary potential is the simple axion cosine potential: $V(\phi) =\Lambda^4 \left[ 1+\cos(\phi/F_a)\right]$. Natural inflation yields power-law spectra for the primordial fluctuations.  The parameter space of natural inflation is constrained by the Planck observations \cite{Akrami:2018odb}. In the allowed region, it predicts primordial tensor fluctuations large enough to be within reach of forthcoming CMB experiments. Axion monodromy \cite{Silverstein:2008sg} is instead a string-motivated UV completion of axion inflation. In this model, the primordial power-law spectrum typical of slow-roll inflation is modulated by oscillatory features. Planck data are consistent with a smooth power-law spectrum of primordial fluctuations, and do not show statistically significant evidence for the presence of periodic features \cite{Meerburg:2013dla,Akrami:2018odb}.

It is remarkable that cosmological observations can constrain the parameter space of axions and ALPs in many different ways, and that this kind of particles might be related to several of the outstanding open problems in cosmology that currently point to new physics. Future cosmological data will definitely allow to shed some light on these issues. On the other hand, axion cosmology will also benefit from theoretical efforts aimed at advancing our understanding of the topological properties of QCD in the high temperature regime.

\section{Experiments searching for axions and axion-like particles}
\label{sec:experiments}

This section gives a brief overview on experimental concepts to search for the QCD axion and axion-like particles (ALPs), thereby focusing on approaches to exploit couplings of such particles to photons. 
Exemplary three larger scale experiments under construction or planned for at DESY in Hamburg are coarsely described.
For a more  detailed overview on the experimental landscape see for example \cite{Irastorza:2018dyq} or \cite{Graham:2015ouw}.
In the following, the term axion will be used as a synonym for axions and ALPs if not stated otherwise.

Lightweight scalar or pseudoscalar axions decay to two photons, however with lifetimes many orders of magnitude larger than the age of the universe so that it is hard to exploit this effect experimentally.  
More promising is to base experiments on the related oscillation of photons into axions and vice versa in presence of a background magnetic field as first proposed by P.\,Sikivie \cite{Sikivie:1983ip}:
axions passing a magnetic field might convert into photons, photons shone into a magnetic field (perpendicular to their momentum) might convert to axions. 
Three different concepts rely on this effect. They are sketched in the following sections.

\subsection*{Haloscopes}
look directly for the dark matter constituents of our milky way. A dark matter axion entering a magnetized volume might convert into photon. In the milky way halo dark matter axions move at non-relativistic speeds so that the photon energy is given by the axion rest mass with an O(10$^{-6}$) correction. 
The power P generated by dark matter axions in an experiment is given by
\begin{equation}
P = \eta g_{a \gamma \gamma}^2 \left( \frac{\rho_a}{m_a} \right) B^2 V \, PB
\end{equation}
with the experimental parameters $\eta$ (efficiency), $B$ (magnetic field strength), $V$ (volume) and $PB$ (power built-up factor of the resonant amplification). 
Nature is providing $g_{a \gamma \gamma}$ (axion-photon coupling), $\rho_a$ (local dark matter axion number density) and $m_a$ (axion mass).
Assuming standard dark matter halo models and the QCD axion making up all the dark matter, typical expectations are $P\approx10^{-22}\,W$ making such experiments extremely challenging. 
Nevertheless, the ADMX experiment has now reached a sensitivity to probe for such dark matter in the $\mu eV$ mass region \cite{Du:2018uak} (cfr. Fig.~\ref{fig:haloscopes}).
ADMX is a prime example for exploiting resonant microwave cavities for axion dark matter searches. 
However, tuning cavities with a relative linewidth of about $10^{-5}$ to candidate axion masses spanning two to three orders of magnitude remains challenging. 
The experiment MADMAX \cite{Brun:2019lyf} strives for an alternative approach with a movable booster of dielectric disks embedded in a high field dipole magnet.
MADMAX will be sited at DESY in Hamburg and could be installed by 2026.
	
\subsection*{Helioscopes}
try to detect axions produced in the solar core. Such particles have energies related to the temperatures in the solar center and are highly relativistic. They can convert to X-rays when passing a transverse magnetic field. 
Therefore helioscopes basically consist of dipole or toroidal magnets tracking the sun and looking for X-ray photons being generated in the light-tight magnet bores. 
Their typical mass reach extends from 0 up to about 1\,eV.
The probability to observe an axion-to-photon conversion is given by  
\begin{equation}
P_{a \rightarrow \gamma} = A g_{a \gamma \gamma}^2 B^2 \left( \frac{\sin(qL/2)}{q} \right)^2  \ \ 
with \ q=\frac{m_a^2}{2E}
\end{equation}
with A being the aperture area of the magnet bore, L the length of the magnetic field B and E the energy of the solar axion. For $q\ll1$ the equation reduces to
\begin{equation}
P_{a \rightarrow \gamma} = A g_{a \gamma \gamma}^2 B^2 L^2 / 4
\end{equation}
It should be noted that helioscopes are also sensitive to axion produced in the sun via their couplings to electrons (see Fig.\,\ref{fig:sensitivity})
The CAST experiment at CERN has performed the most sensitive search for solar axions at present 
\cite{Anastassopoulos:2017ftl}. \medskip \\
The International Axion Observatory (IAXO) \cite{Armengaud:2014gea} is the successor of CAST being able to check for the astrophysical hints mentioned above and to explore the QCD axion in the mass range around 1\,meV, which is not accessible by any other technology.
IAXO will increase the sensitivity on $g_{a \gamma \gamma}$ by roughly a factor 20 compared to CAST. \\
A first step will be the BabyIAXO prototype to mainly test the magnet concept, new X-ray optics and detectors. However, it will also have a physics reach more than a factor 3 beyond CAST and could be ready in 2024.
DESY in Hamburg is the potential host for BabyIAXO and IAXO. 
	
\subsection*{Light-shining-through-walls}
experiments aim for producing and detecting axions in the laboratory. They do not rely on cosmological or astrophysical assumptions. 
In the first section of such an experiment, laser light is shone through a strong magnetic field, where axions can be generated via a reverse process sketched above. A second section of the experiment is separated from the first one by a light-tight wall which can only be surpassed by axions. These particles would stream through a strong magnetic field behind the wall allowing for a re-conversion into photons. This effect will give the impression of light-shining-through-a-wall (LSW).
The probability for a photon-axion-photon conversion is very similar to the helioscope case, but includes now also the axion generation part. For $q\ll1$:
\begin{equation}
P_{\gamma \rightarrow a \rightarrow \gamma} = \frac{1}{16}  g_{a \gamma \gamma}^4 B^4 L^4 
\label{eq:lsw}
\end{equation}
for a symmetric set-up with equal magnetic field strength B and length L before and behind the wall.\\
LSW experiments have probed for axions in a model independent fashion \cite{ Ehret:2010mh, Betz:2013dza,Ballou:2015cka}, but with sensitivities for $g_{a \gamma \gamma}$ about three orders of magnitude less than helioscopes. 
ALPS\,II presently under construction at DESY in Hamburg, based on 20 dipole magnets from the former HERA proton accelerator, aims for going beyond the CAST sensitivity for axion masses below 0.1\,meV by incorporate mode-matched optical resonators before and behind the wall \cite{Bahre:2013ywa}. 
Equation\,(\ref{eq:lsw}) at ALPS\,II will be modified to 
\begin{equation}
P_{\gamma \rightarrow a \rightarrow \gamma} = \frac{1}{16}  PB_p PB_r g_{a \gamma \gamma}^4 B^4 L^4 
\end{equation}
with $PB_p$ and $PB_r$ denoting the power built-up factors of the resonators before and behind the wall. 
ALPS\,II is aiming for $PB_p=5,000$ and $PB_r=40,000$.
This will allow to probe for the astrophysical hints mentioned previously. Data taking is scheduled to start in 2020.

\subsection*{Summary}
Different experimental approaches to find axions are being pursued. 
An overview on experiments exploiting axion-photon couplings is summarized in Table\,\ref{tab:experiments}.
Only few experiments are taking data at present, but there will be many more results available in the next decade as partly shown in the figures \ref{fig:gag_ma} and \ref{fig:haloscopes}.
\begin{table}[htb]
	\centering
		\begin{tabular}{| l | c | c | c |} 
		\hline
			\bf{Axion property} & \bf{Haloscope} & \bf{Helioscope} & \bf{LSW}  \\ \hline
			\bf{Spin-parity} & yes & perhaps & yes \\ 
			\bf{Mass} & yes & perhaps & perhaps \\
			$\mathbf{ g_{a \gamma \gamma}}$ & no & no & yes \\
			$\mathbf{g_{a \gamma \gamma} \times flux}$ & yes & yes & -- \\
			\bf{Electron coupling} & no & yes & no \\
			\bf{QCD axion in reach} & yes & yes & no \\
			\bf{Assumptions} & cosmology & solar physics & -- \\ \hline
			\bf{Experiments taking data} & $\approx$5 & 0 & 0\\
			\bf{Experiments under construction} & $\approx$4 & 0 & 1\\
			\bf{Experiments proposed} & $\approx$10 & 2 & 2\\
			\hline
		\end{tabular}
	\caption{A comparison of different approaches to find the axion}
	\label{tab:experiments}
\end{table}

Within the update process of the European strategy on particle physics launched in 2018, the axion community is proposing a ``A European Strategy Towards Finding Axions and Other WISPs'' \cite{espp}.

\end{document}